\documentclass[a4paper]{jpconf}
\usepackage{amssymb}
\usepackage{amsmath}
\usepackage{amssymb}
\usepackage{amsbsy}
\usepackage{bm}
\usepackage{bbold}
\usepackage{wrapfig}
\usepackage{graphicx}
\usepackage{siunitx}
\newcommand{\appropto}{\mathrel{\vcenter{
  \offinterlineskip\halign{\hfil$##$\cr
    \propto\cr\noalign{\kern2pt}\sim\cr\noalign{\kern-2pt}}}}}

\newcommand{\GG}{{\sf G}}
\newcommand{\PP}{{\sf P}}
\newcommand{\TT}{{\sf T}}

\newcommand{\Ss}{{\sf S}}

\newcommand{\be}{\begin{equation}}
\newcommand{\ee}{\end{equation}}
\newcommand{\bea}{\begin{eqnarray}}
\newcommand{\eea}{\end{eqnarray}}

\newcommand{\vv}[2]{ \left( \begin{array}{cc}   #1 \\ #2 \end{array} \right)}

\renewcommand\({\left(}
\renewcommand\){\right)}
\renewcommand\[{\left[}

\begin{document}
\title{MADMAX: A new way of probing QCD Axion Dark Matter with a Dielectric Haloscope -- Foundations}

\author{Stefan Knirck for the MADMAX interest group$^{\small \dagger}$}

\address{Max-Planck-Institute for Physics, Werner-Heisenberg-Institute, Foehringer Ring 6, 80805
Munich, Germany}

\ead{knirck@mpp.mpg.de}

\vspace{0.2cm}

\address{{\bf $^{\dagger}$ The MADMAX interest group:} \\{\bf MPI for Radioastronomy} - Bonn, Germany: M.\,Kramer, G.\,Wieching; \\{\bf DESY Hamburg} - Hamburg, Germany: H.\,Kr\"uger, A.\,Lindner, C.\,Martens, J.\,Schaffran; \\{\bf University of Hamburg} - Hamburg, Germany: E.\,Garutti, A.\,Schmidt; \\{\bf MPI for Physics} - Munich, Germany: A. Caldwell, G.\,Dvali, C.\,Gooch, A.\,Hambarzumjan, S.\,Knirck, B.\,Majorovits, A.\,Millar, G.\,Raffelt, O.\,Reimann, F.\,Steffen; \\{\bf CEA IRFU} - Saclay, France: P. Brun, L. Chevalier; \\{\bf University of Zaragoza} - Zaragoza, Spain: J. Redondo}

\begin{abstract}
In contrast to WIMPs, light Dark Matter candidates have increasingly come under the focus of scientific interest. In particular the QCD axion is also able to solve other fundamental problems such as CP-conservation in strong interactions.
Galactic axions, axion-like particles and hidden photons can be converted to photons at boundaries between materials of different dielectric constants under a strong magnetic field. Combining many such surfaces, one can enhance this conversion significantly using constructive interference and resonances. The proposed MADMAX setup containing 80 high dielectric disks in a \SI{10}{\tesla} magnetic field would probe the well-motivated mass range of $40$--\SI{400}{\micro\electronvolt}, a range which is at present inaccessible by existing cavity searches. We present the foundations of this approach and its expected sensitivity.
\end{abstract}

\section{Introduction}
Axions are prime cold dark matter (DM) candidates, originally proposed to solve the strong CP-problem. This originates from the fact that quantum chromodynamics (QCD) allows for CP-violation via a term $\mathcal{L} \propto \theta \, G^a_{\mu \nu} \tilde{G}^{\mu \nu}_a$, where $\theta$ is an angular natural constant and $G^a_{\mu \nu}, \tilde{G}^{\mu \nu}_a$ are the gluon field strength tensor and its dual. However, measurements of the neutron electric dipole moment constrain $|\theta| < \SI{e-10}{}$, which appears unnaturally small. The axion is introduced by promoting $\theta$ to a field whose potential is minimized at $\theta = 0$. Starting from a non-zero value $\theta_I$ it will roll down to eventually coherently oscillate around the CP-conserving minimum $\theta = 0$, naturally explaining the smallness of $\theta$ -- and giving rise to a new, weakly interacting, light particle: the axion \cite{Peccei:2006as}.

Causality implies that at some early time our universe consisted of uncorrelated patches with different initial angles $\theta_I$.
If inflation happened after this time (Scenario A), then our visible universe was inflated out of one of these patches, leaving the axion mass broadly unconstrained. If inflation happened before this time (Scenario B), these patches give rise to axion strings and domain walls, constraining the axion mass to $\SI{50}{\micro\electronvolt} \lesssim m_a \lesssim \SI{200}{\micro\electronvolt}$, cf. e.g. \cite{Kawasaki:2014sqa}.

Various different experiments have been proposed or are underway to search for axion DM, mostly relying on the interaction between axions and photons; for a review cf. e.g. \cite{Graham:2015ouw}.
 In this note we briefly review the foundations of axion DM detection with a dielectric haloscope like MADMAX, as recently introduced in \cite{TheMADMAXWorkingGroup:2016hpc,theoretical-foundations}.

\section{Axion Electrodynamics}
The interactions of electromagnetic (EM) fields with axions are described by the Lagrangian

\begin{equation}\label{lagrangian}
{\cal L} = -\frac{1}{4}F_{\mu\nu}F^{\mu\nu}-J^\mu A_\mu+\frac{1}{2}\partial_\mu a \partial^\mu a
-\frac{1}{2}m_a^2a^2 -\frac{g_{a\gamma}}{4}F_{\mu\nu}\widetilde F^{\mu\nu}a ,
\end{equation}
where $F_{\mu\nu}$ is the EM field strength tensor, $A^\mu$ the respective vector potential, $J^\mu$ the EM 4-current. $a$ denotes the axion field, $m_a$ its mass, and $g_{a\gamma}$ a
coupling constant of dimension (energy)$^{-1}$. We use natural units $\hbar = c = 1$ and Lorentz-Heaviside convention with the fine-structure constant $\alpha = e^2/4\pi$ with $e$ the elementary charge. For QCD axions $g_{a\gamma}$ is proportional to the axion mass and we expect~\cite{diCortona:2015ldu}
\begin{equation}
g_{a\gamma}=-\frac{\alpha}{2\pi f_a}\,C_{a\gamma}
=-2.04(3)\times10^{-16}~{\rm GeV}^{-1}\,\left(\frac{m_a}{1\,\mu{\rm eV}}\right)\,C_{a\gamma}\,,\label{eq:gag}
\end{equation}
with $f_a$ the axion decay constant anti-proportional to $m_a$ and $C_{a\gamma}$ a model dependent constant of order one. For more general models (axion-like-particles) $g_{a\gamma}$ is a free parameter.

If axions make up all DM, then their local abundance would be ${{\sim \SI{0.3}{\giga\electronvolt\per\cubic\centi\metre}} / m_a}$. With ${m_a \sim \SI{100}{\micro\electronvolt}}$ this corresponds to a number density of ${\sim \SI{3e12}{\per\cubic\centi\metre}}$, while their {de Broglie wavelength} is ${\lambda_{\rm DB}~=~2\pi/(m_a v_a) \sim \SI{10}{\metre}}$ due to their non-relativistic velocity ${v_a \sim \mathcal{O}(10^{-3})}$ and small mass. Therefore it is sufficient to treat DM axions as a classical field and we may just solve the classical equations of motion following from Eq.~\eqref{lagrangian}. 
Further assuming a static, external magnetic field $\mathbf{B}_{\rm e}$, this leads to modified inhomogeneous Maxwell equations: \cite{theoretical-foundations}
\begin{subequations}
\begin{eqnarray}
\epsilon \bm\nabla\cdot {\bf E} 
&=&  \rho
- g_{a\gamma} {\bf B}_{\rm e}\cdot{\bm\nabla} a\,,
\label{eq:Maxwell-aa}\\
{\bm\nabla}{\bm \times} {\bf H}-\epsilon\dot{\bf E}
&=&{\bf J}+g_{a\gamma}{\bf B}_{\rm e} \dot a\,,\label{eq:Maxwell-bb}
\\
\ddot a-{\bm\nabla}^2 a +m_a^2 a &=&
g_{a\gamma}{\bf E}\cdot{\bf B}_{\rm e}\,,\label{eq:Maxwell-cc}
\end{eqnarray} 
\end{subequations}%
where $\rho$ and ${\bf J}$ are the electric charge density and current respectively, ${\bf E}$ and ${\bf B}$ the electric and magnetic field, $\epsilon$ the dielectric constant and ${\bf H}$ the macroscopic magnetic field. Here ${\bf H}$ is without the contribution of ${\bf B}_{\rm e}$ and ${\bf J}$ respectively without the current needed to generate ${\bf B}_{\rm e}$.
Note, this implies that the oscillating axion field $a \propto \exp(-i m_a t)$, induces an electric field
\begin{equation}
\label{eq:Ea}
{\bf E}_a(t)=-\frac{g_{a\gamma}{\bf B}_{\rm e}}{\epsilon}\,a(t) ~~\sim~~ 1.3\times10^{-12}~{\rm V}/{\rm m}~\frac{B_{\rm e}}{10~{\rm T}}~
\frac{|C_{a\gamma}|f_{\rm DM}^{1/2}}{\epsilon},
\end{equation}
where $f_{\rm DM}$ is the fraction of DM made up by axions.

Further, the homogeneous Maxwell equations are unchanged. Thus, the usual EM boundary conditions for ${E}$ and ${H}$ fields remain, i.e., on boundaries of two media with different dielectric constant $\epsilon$ their component parallel to the boundary must be continuous. However, Eq.~\eqref{eq:Ea} suggests a jump of the induced electric field on such a boundary. In order to fulfill all Maxwell equations, EM waves must be emitted from the boundary compensating this discontinuity \cite{Horns:2012jf}. Their amplitudes are
\begin{eqnarray}
  {E}_{1}^\gamma=+\Delta E^a\,
  \frac{\epsilon_2n_1}{\epsilon_1n_2+\epsilon_2n_1}\,,
%\\[1ex]
~~~
  {E}_{2}^\gamma=-\Delta E^a\,
  \frac{\epsilon_1n_2}{\epsilon_1n_2+\epsilon_2n_1}\,,
%  \\[1ex]
~~~
  {H}_{1,2}^\gamma=-\Delta E^a\,
  \frac{\epsilon_1\epsilon_2}{\epsilon_1n_2+\epsilon_2n_1}\,,
\end{eqnarray}
with $\Delta E^a = \left({E}_{2}^a-{E}_{1}^a\right)$ and $n_i = \sqrt{\epsilon_i}$ the refractive indices of the media.
This leads to an emitted power from the boundary of
\begin{equation}
{P_\gamma}
=2.2\times 10^{-27}\,\si{\watt}
\(\frac{A}{\SI{1}{\square\metre}}\) \(\frac{B_{\rm e}}{10~{\rm T}}\)^2 f(\epsilon_1, \epsilon_2) \, C_{a\gamma}^2 \, f_{\rm DM}\,.
\end{equation}
at the axion oscillation frequency $m_a$ with a dispersion from the cold dark matter velocity ${\sim \mathcal{O}(\SI{e-6}{}) m_a}$. Here $A$ is the interface area between the dielectrics and typically ${f(\epsilon_1, \epsilon_2) \leq 1}$, depending on the dielectrics. 

%\newpage
\section{Power Boost}
\begin{wrapfigure}[19]{r}{0.52\textwidth} 
\vspace{-2.7em}
\includegraphics[width=0.52\textwidth]{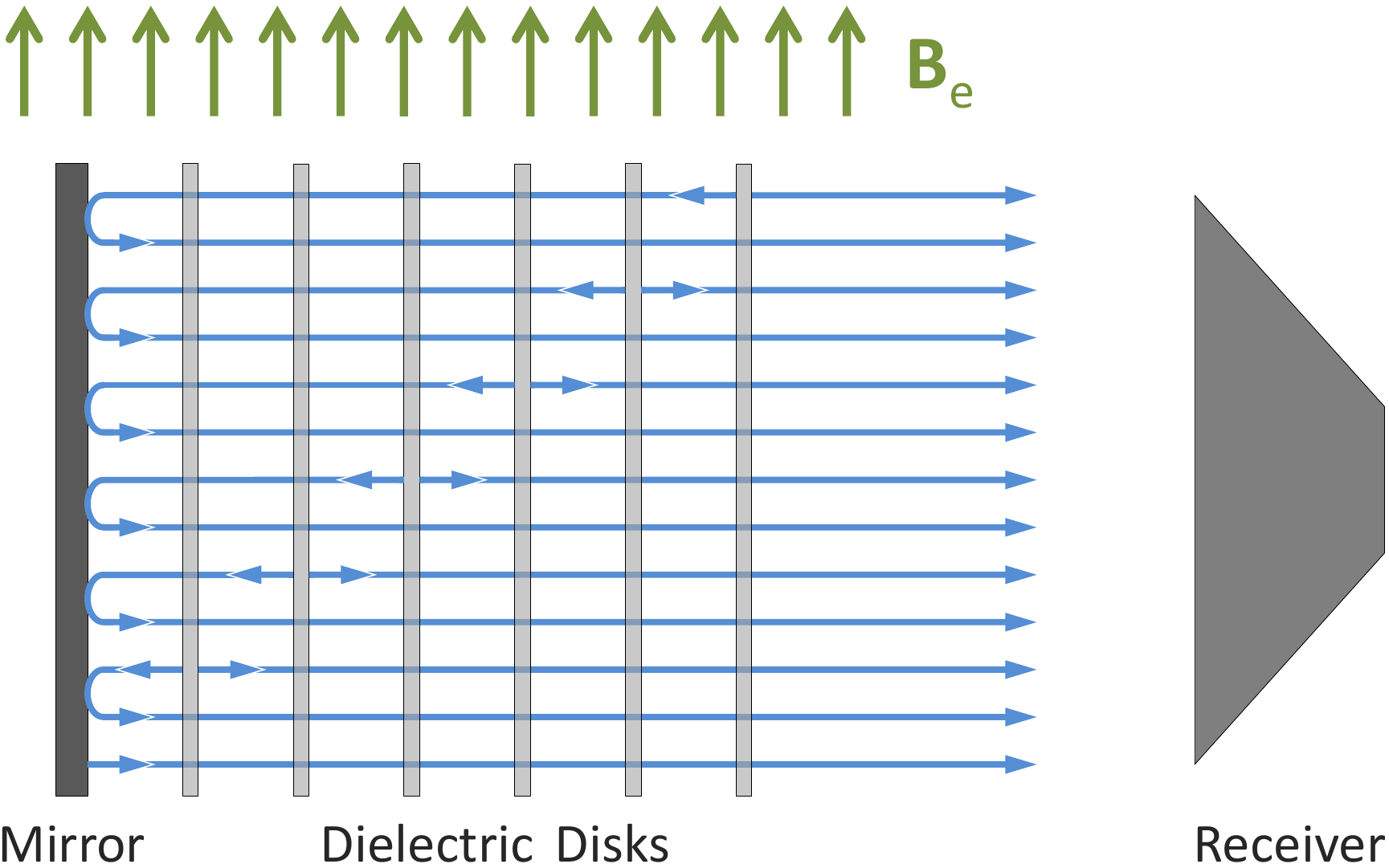}
\caption{MADMAX idea: Each interface coherently emits EM radiation due to the presence of the axion field. Depending on the axion mass and the disk placements these may interfere constructively and 'boost' the emitted power compared to a single mirror setup by 4--5 orders of magnitude.}
\label{fig:interface}
\end{wrapfigure}
By employing many dielectric disks in a strong magnetic field $\mathbf{B}_{\rm e}$, each surface emits this radiation coherently, since the axion coherence length guaranteed by its de Broglie wavelength is greater than the size of the setup. By placing the disks accordingly those contributions may interfere constructively while also internal reflections may resonantly enhance the emission, both leading to a total emission much higher than what would be expected from a single surface. We define the power boost factor $\beta^2$ as the ratio between the power emitted by such a `booster' and the power emitted by a single mirror.

\begin{figure}[t]	
	\begin{minipage}[b]{0.5\textwidth}
		\includegraphics[width=\textwidth]{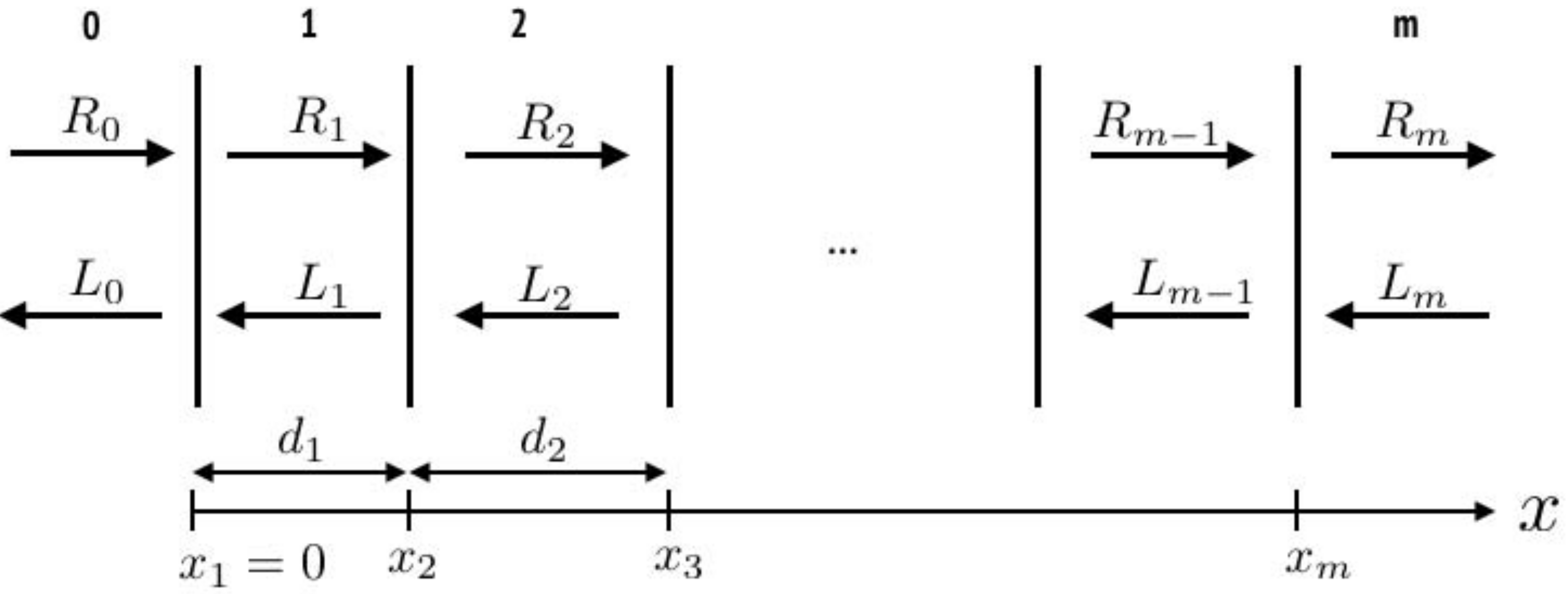}
	\end{minipage}\hfill%
	\begin{minipage}[b]{0.48\textwidth}\caption{\label{fig:scheme}Parametrization of left- and right-propagating waves with amplitudes $L_r$ and $R_r$ in our system of different regions with different dielectric constants $\epsilon_r$ and thickness $d_r$.}
	\end{minipage}
\end{figure}

The fields of an EM wave emitted by such a system can be calculated by identifying the left and right-propagating EM field amplitudes in each region $r$ with dielectric constant $\epsilon_r$ according to Fig.~\ref{fig:scheme}
by a vector $(R_r, L_r)^T$ and defining the transfer matrices
\begin{eqnarray}
\text{{Reflection / Transmission:}}~~~~~~~&\text{{Phase Propagation:}~~}&\text{{Axion Induced Fields:}}~~~~~~~~~~~~ \nonumber \\[1ex]
\GG_r\!=\!\frac{1}{2n_{r+1}}\(\begin{array}{cc}
n_{r+1}{+}n_r & n_{r+1}{-}n_r\\
n_{r+1}{-}n_r & n_{r+1}{+}n_r \end{array}\),
&\PP_r=\(\begin{array}{cc}
e^{+i \delta_r}\!&\!0\\
0&e^{-i \delta_r} \end{array}\),
&~\Ss_r=\frac{A_{r+1}-A_{r}}{2}\,
\mathbb{1}
\end{eqnarray}
where $\delta_r = \omega n_r d_r$ is the phase depth of region $r$. Then the amplitudes in region $m$ can be expressed in terms of region $m-1$ and recursively in terms of region 0 as 
\begin{eqnarray}
\vv{R_m}{L_m}
&=& \GG_{m-1} \PP_{m-1} \vv{R_{m-1}}{L_{m-1}} + E_0\Ss_{m-1}\vv{1}{1}
=\ldots
\nonumber\\[1ex]
&=& \TT^m_0 \vv{R_{0}}{L_{0}}
+E_0\sum_{s=1}^m  \TT^{r}_s \Ss_{s-1}\vv{1}{1}\,, \label{transfereq}
\end{eqnarray}
with a transfer matrix from region $a$ to $b$ with $b<a$
\begin{equation}
\TT^a_b = \GG_{a-1}\PP_{a-1}\GG_{a-2}\PP_{a-2}\,\ldots\, \GG_{b+1}\PP_{b+1}\GG_{b}\PP_{b} , \quad \TT^a_a=\mathbb{1}. \label{eq:emtransfer}
\end{equation}

The power boost factor is then easily obtained by the rightmost outgoing amplitude as $\beta^2 = |R_m/E_0|^2$, requiring $R_0 = L_m = 0$. Similarly the reflectivity of the system is just given by $\mathcal{R} = R_m / L_m$, requiring $R_0 = 0$.
As an example, it is straightforward to calculate the reflectivity of a single disk, given by
\begin{equation}
	{\cal R}_{\rm disk}=\frac{(n^2-1)\,\sin\delta}{i\,2n\cos\delta+(n^2+1)\,\sin\delta}\,,
\end{equation}
where $n$ is its refractive index and $\delta = \omega n d$ its optical thickness as defined above. For thicknesses of $\delta=0,\pi,2\pi,\ldots$ the disk does not reflect any radiation, i.e., is transparent; while for  $\delta=\pi/2,3\pi/2,\ldots$ the disk becomes maximally reflective. In a many disk system one can for example place the disks at distances to coherently add up the signal for frequencies at which they are transparent or make the system maximally resonant at which they are maximally reflective. Combining both effects enables to form broadband boost factors as described below.

We remark that the classical results can be confirmed by an explicit quantum-field calculation. Evaluating the matrix element for the axion-photon conversion under an external magnetic field ${\cal M}=\langle {\rm f}|H_{a\gamma}|{\rm i}\rangle={g_{a\gamma}}/{2\omega V}\int d^3{\bf r}\,
 e^{i{\bf p}\cdot{\bf r}}\,{\bf B}_{\rm e}({\bf r})\cdot{\bf E}^*_{\bf k}({\bf r})$
and using non-momentum-eigenstates and non-translational invariant Garibian wave functions as out-states for the emitted photons, one may derive the same results than with a classical calculation \cite{qft-calculation}.

\section{Power Boost Properties}
\begin{wrapfigure}[17]{l}{0.46\textwidth} 
	\centering
	\vspace{-1em}
	\includegraphics[width=0.45\textwidth]{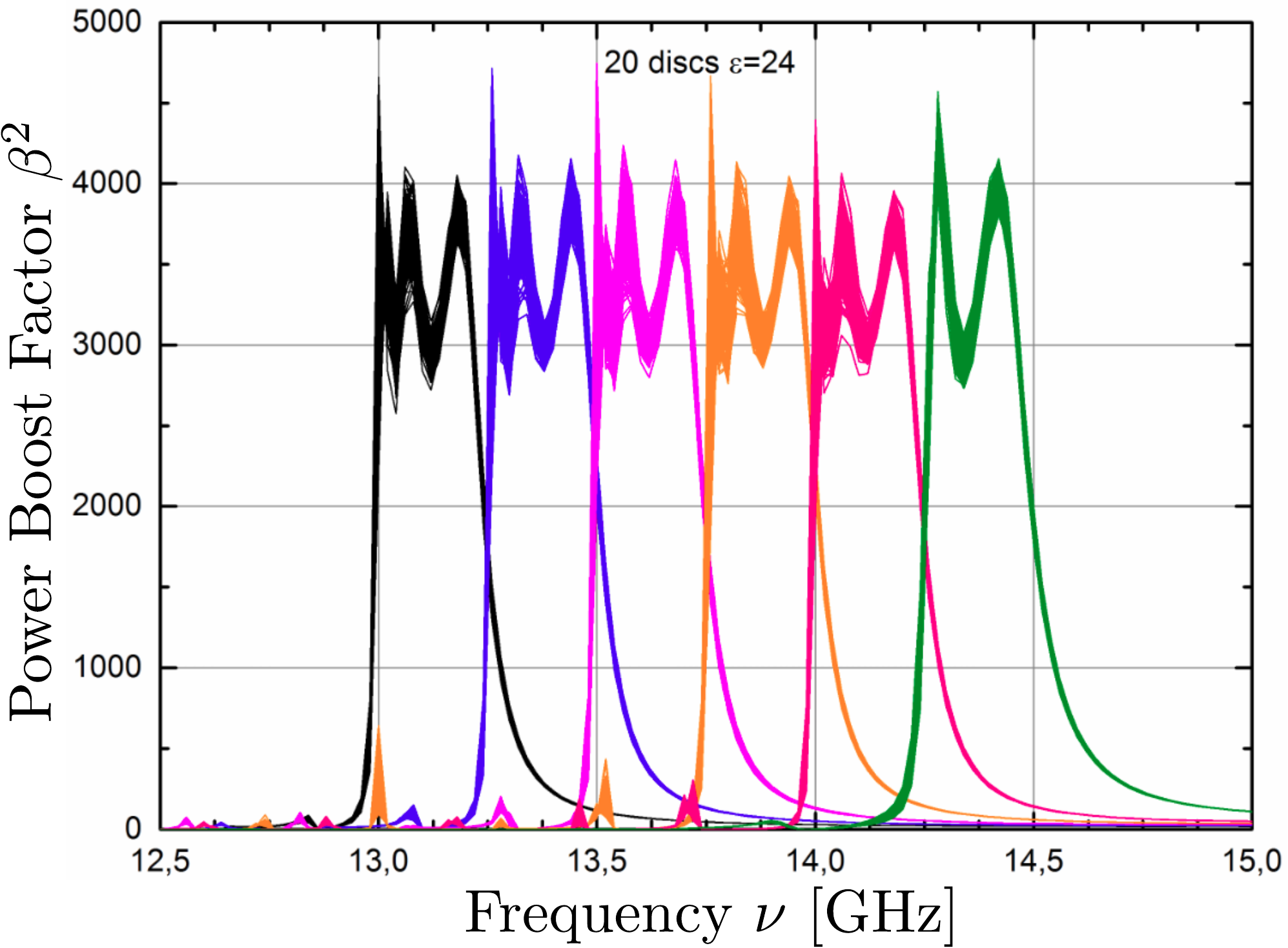}
	\vspace{-0.7em}
	\caption{Exemplary broadband boost factors for 20 disks, $\epsilon = 24$ and $d = \SI{1}{\milli\metre}$. The frequency response can be changed with the disk positions.}
	\label{fig:bf-bandwidths}
\end{wrapfigure}

\begin{figure}[t]
	\includegraphics[height=10.7em]{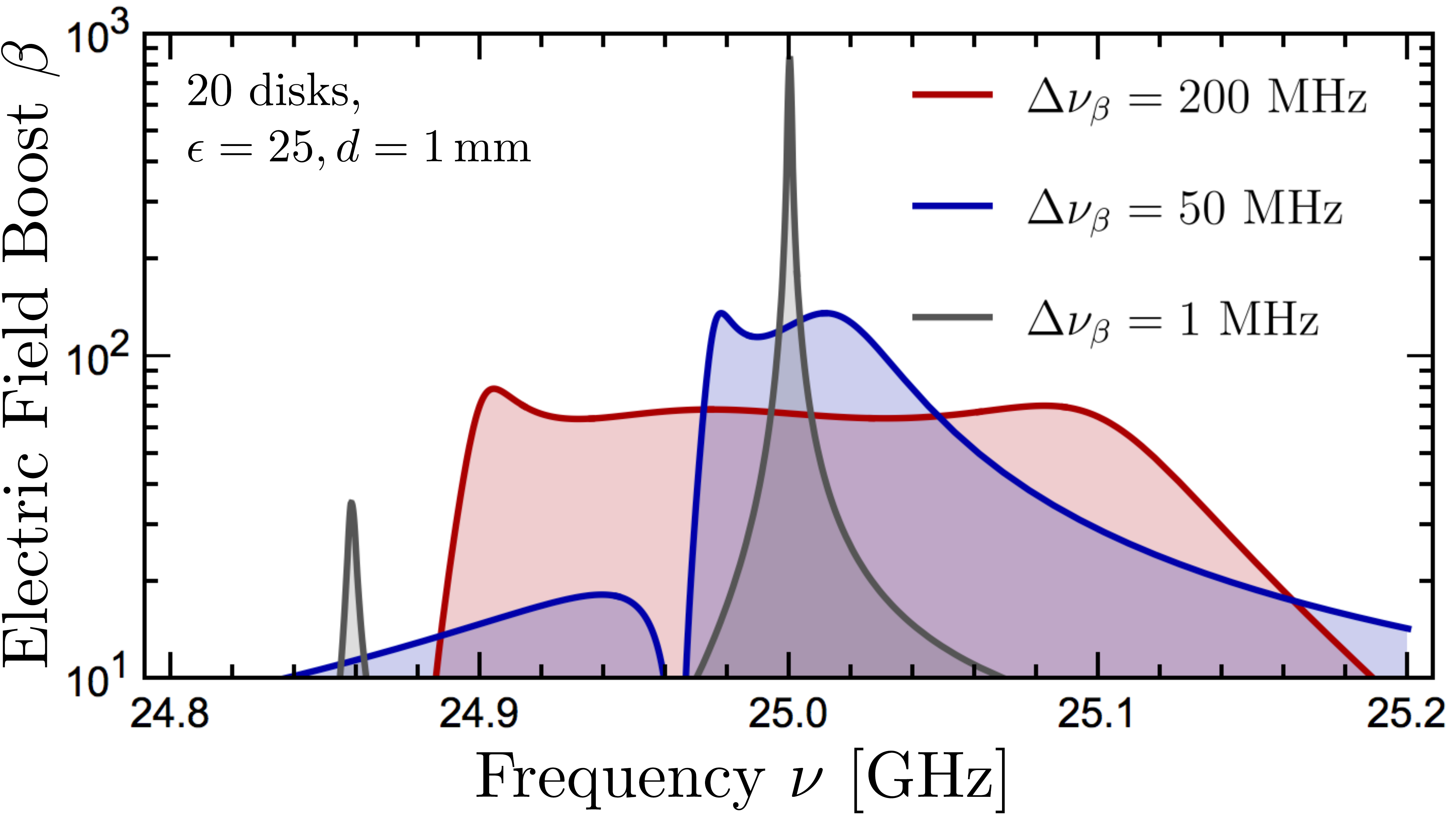}
	\hfill
	\includegraphics[height=10.7em]{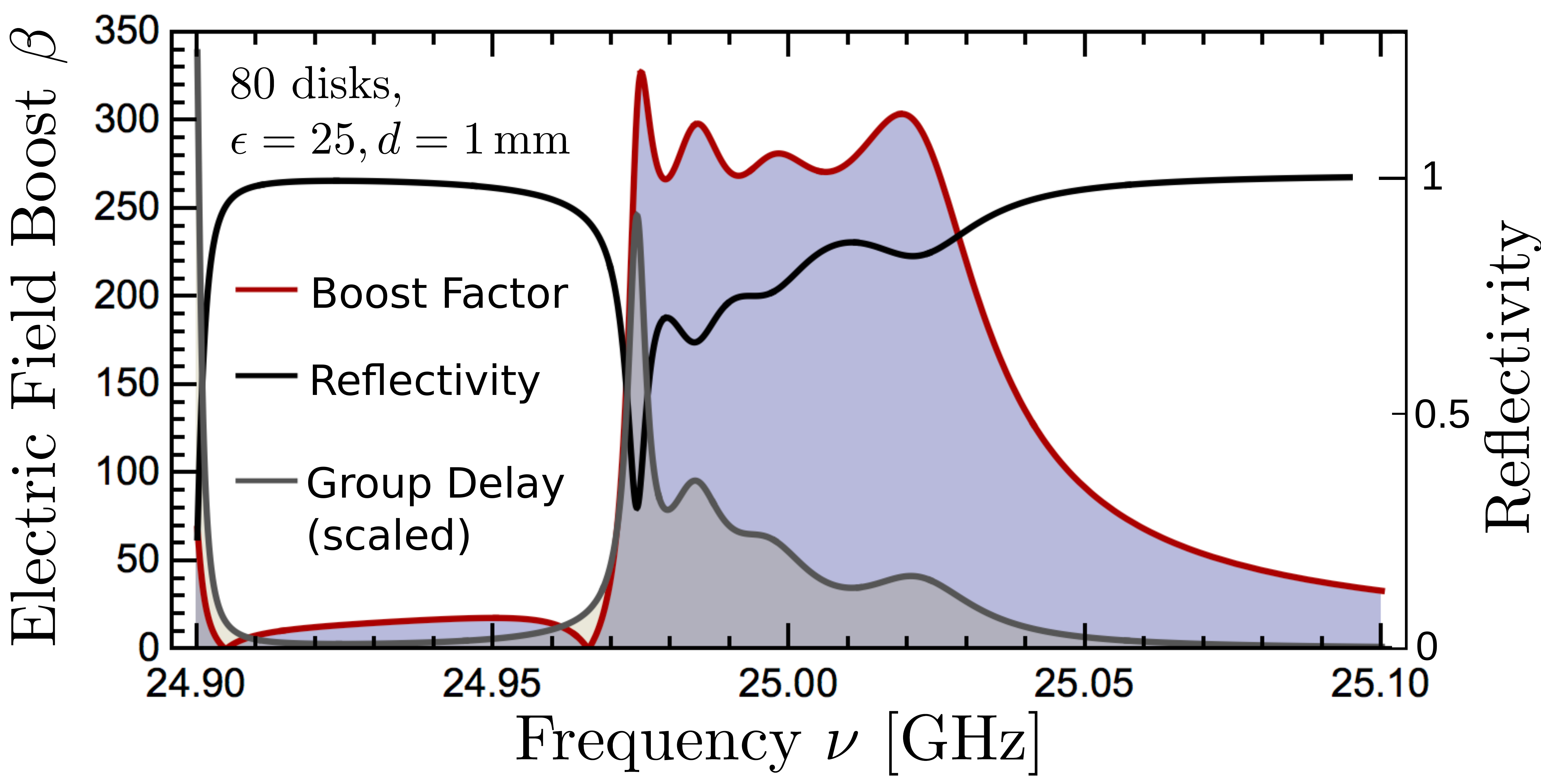}
	\caption{Boost Factor Properties. \textbf{Left:} Area Law. $\int \beta^2 d \nu$ is roughly constant for all spacing configurations. \textbf{Right:} The boost factor is correlated with measurable quantities such as the reflectivity of the total disk system or the group delay of a reflected signal. Adapted from \cite{TheMADMAXWorkingGroup:2016hpc,Millar:2017eoc}.}
	\label{fig:bf-properties}
\end{figure}

By optimizing the dielectric disk distances the frequency response of the boost factor can be controlled. Generally, the average distance approximately sets the frequency while their variance amounts for broadening the bandwidth. For a broadband scan over various frequencies, i.e., axion masses $m_a$, a top-hat response is ideal. By numerically optimizing the disk positions, one can find such solutions approximately, as exemplary shown Fig.~\ref{fig:bf-bandwidths}.
Crucially, the area under the power boost factor curve $\int \beta^2 d \nu$ is conserved (Area~Law). For an explicit proof cf.~\cite{theoretical-foundations}. While for integrating over ${0~\leq~\nu~\leq~\infty}$ this holds exactly, it is still a good approximation in the region containing the main peak.  Therefore one can trade bandwidth with power boost by changing disk positions, but not simultaneously gain in both, cf. Fig.~\ref{fig:bf-properties},~left.
After scanning a broadband frequency range, one may therefore specifically perform rescans of possible detection candidates with a much larger but narrower boost factor. 
The boost factor is correlated to measurable quantities such as the reflectivity of the disk system, as shown in Fig.~\ref{fig:bf-properties},~right.
This correlation can be understood for example, by realizing that both quantities can be calculated using the same transfer matrix formalism.
Since measuring phase instead of magnitude may involve less experimental systematics, one may also consider the group delay of a reflected signal, which carries the same information. 

\newpage
\section{Conclusion and Outlook}
\begin{wrapfigure}[22]{r}{0.44\textwidth}
	\vspace{-4em}
	\centering
	\includegraphics[width=0.44\textwidth]{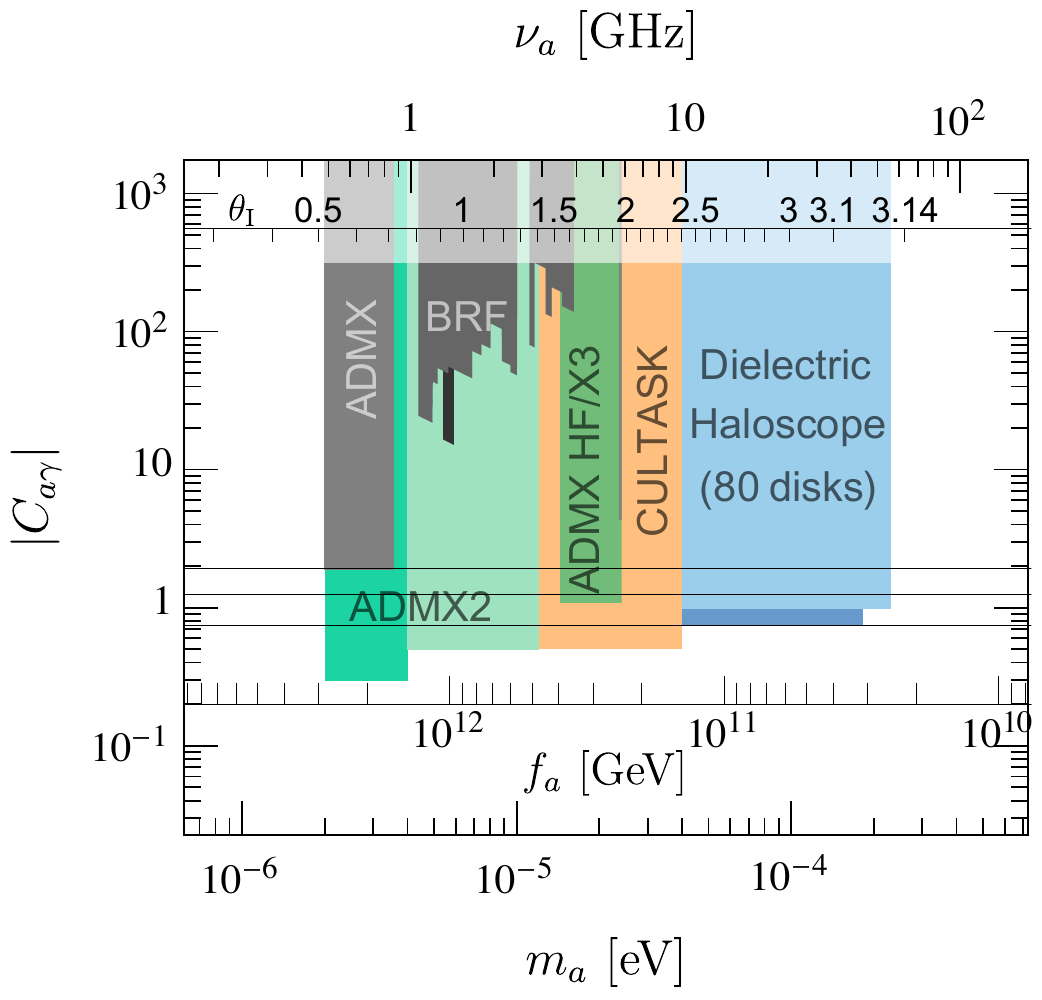}
	\vspace{-1.5em}
	\caption{Discovery potential of the proposed 80 disk MADMAX setup, achievable within a scanning campaign of 5 years~(dark~blue) and additional {2~years~(light~blue)}. Other colors show other experimental efforts, cf.~e.g.~\cite{Graham:2015ouw}, to which MADMAX is broadly complementary, probing masses corresponding to initial values of $\theta_I \gtrsim 2.4$ in Scenario~A and the preferred mass range in Scenario~B.}
	\label{fig:discovery-potential}
\end{wrapfigure}
We have reviewed how the presence of axion dark matter may cause emission of EM waves from a dielectric interface in presence of a magnetic field and how a dielectric haloscope consisting of many dielectric disks may boost this power.

This approach is particularly appealing for masses $\gtrsim \SI{40}{\micro\electronvolt}$, where traditional resonant cavity searches struggle to achieve the required volume. Assuming 80 dielectric disks and sufficiently low losses 
(${d=\SI{1}{\milli\meter}, A=\SI{1}{\square\meter}, \epsilon=25, \tan \delta \sim \num{e-5}}$) a boost factor of ${\sim\SI{5e4}{}}$ can be achieved over a bandwidth of \SI{50}{\mega\hertz}. For $B_{\rm e} = \SI{10}{\tesla}$ this leads to a signal of ${\sim \SI{e-22}{\watt}}$, measurable with state-of-the-art cryogenic detectors. Assuming a system noise temperature of $T_{\rm sys} = \SI{8}{\kelvin}$, one could scan the range of ${\SI{40}{\micro\electronvolt} \lesssim m_a \lesssim \SI{120}{\micro\electronvolt}}$ with a sensitivity of up to $|C_{a \gamma}| = 0.75$ within 5 years. Using quantum-limited detectors one could extend the search to masses ${m_a \lesssim \SI{230}{\micro\electronvolt}}$, as outlined in Fig.~\ref{fig:discovery-potential}. Note, that this covers the preferred region in Scenario B described above of ${\SI{50}{\micro\electronvolt} \lesssim m_a \lesssim \SI{200}{\micro\electronvolt}}$. First encouraging R\&D activities are already underway, for more details cf.~\cite{TheMADMAXWorkingGroup:2016hpc,Majorovits:2017xxx}.

\end{document}